\documentclass[
 reprint,
 runinaddress,
 amsmath,amssymb,
 aps,
]{revtex4-2}
\usepackage{graphicx}
\usepackage{dcolumn}
\usepackage{bm}

\usepackage[separate-uncertainty = true]{siunitx}
\DeclareSIUnit{\molar}{M}

\begin{document}

\preprint{APS/123-QED}

\title{Quantitative Stain Mapping in X-ray Virtual Histology}

\author{Dominik John$^{1,2,3}$, David M. Paganin$^{2}$, Marie-Christine Zdora$^{2}$, Lisa Marie Petzold$^{1,4}$, Patrick Ilg$^{1,4}$,\\ Junan Chen$^{1,5}$, Sara Baggio$^{1}$, Johannes B. Thalhammer$^{1,2,4}$, Sami Wirtensohn$^{1,3}$, Julian Moosmann$^{3}$, Jörg U. Hammel$^{3}$, Felix Beckmann$^{3}$, Samantha J. Alloo$^{2}$, Jannis Ahlers$^{2}$, Madleen Busse$^{6}$, Julia Herzen$^{1}$, and Kaye S. Morgan$^{2}$}

\affiliation{\parbox{0.8\textwidth}{%
$^{1}$Research Group Biomedical Imaging Physics, Department of Physics,\\ TUM School of Natural Sciences \& Munich Institute of Biomedical Engineering,\\Technical University of Munich, Garching, Germany\\
$^{2}$School of Physics and Astronomy, Monash University, Victoria, Australia\\
$^{3}$Helmholtz-Zentrum Hereon, Institute of Materials Physics, Geesthacht, Germany\\
$^{4}$Chair of Biomedical Physics, TUM School of Natural Sciences,\\ \phantom{$^{5}$}Technical University of Munich, Garching, Germany\\
$^{5}$ImFusion GmbH, Munich, Germany\\
$^{6}$Department Biological Safety, German Federal Institute of Risk Assessment (BfR), Berlin, Germany
}}
\begin{abstract}
Virtual histology is an emerging field in biomedicine that enables three-dimensional tissue visualization using X-ray micro-computed tomography. However, the method still lacks the specificity of conventional histology, in which parts of the tissue are selectively highlighted using targeted stains. Though some first X-ray stains have been developed to address this issue, their precise location and quantity inside the tissue volume remain largely unknown. In this work, we present a novel approach to virtual histology that simultaneously captures electron number density and X-ray attenuation values through modulation-based X-ray imaging with a structured phase modulator. These complementary measurements enable decomposition of tissue volumes into basis materials, which allows the extraction of three-dimensional maps of molar contrast agent distribution alongside morphological details on the micrometer scale---here demonstrated on murine kidneys. The concentration values are validated against the established method of K-edge subtraction imaging. We also create a direct bridge from X-ray to visible light imaging by detecting the same stain both using classical histology and our proposed X-ray approach. Our methodology opens new possibilities for biomedical research into disease progression by providing quantitative three-dimensional stain mapping across entire tissue volumes alongside high-contrast morphology, enabling deeper insights into disease mechanisms.
\end{abstract}

\maketitle

\section{Introduction}
Histopathology, the microscopic examination of diseased tissue, is the gold standard for diagnosing a multitude of diseases. However, traditional histopathology requires physically cutting tissue into thin slices for examination under an optical microscope, limiting analysis to two-dimensional views that may show histological artifacts \cite{pichat2018survey}.
Though methods for reconstructing three-dimensional information from conventional histological sections have been proposed \cite{kurz20243}, the process is labor-intensive and requires advanced methods for registration due to artifact-induced inconsistencies between slices \cite{paknezhad2020regional}. To gain direct access to three-dimensional information, methods like fluorescent light sheet microscopy \cite{glaser2017light, girkin2018light, freise2023three} and X-ray micro-computed tomography \cite{prajapati2011microct, senter2016role} are therefore active areas of research.

Virtual histology using X-ray micro-computed tomography offers a complementary approach to conventional histology by enabling three-dimensional (3D) tissue examination without physical sectioning or the need for sample transparency, providing new insights into disease mechanisms and tissue architecture \cite{albers2018x, topperwien2018three, zdora2020x, eckermann2021three, massimi2021volumetric, reichardt20213d, savatovic2025high}. 
The acquired 3D volume may also serve as a comprehensive overview of the sample structure, guiding subsequent targeted histological analysis of specific regions of interest \cite{zdora2020x, chen20242d}.

A major challenge for conventional X-ray imaging of soft tissues is their naturally low contrast due to the subtle difference in density across different types of tissue. To address this limitation, various techniques have been developed that exploit not only how X-rays are attenuated by tissue, but also how they are refracted or scattered. These phase-contrast methods include propagation-based imaging \cite{snigirev1995possibilities, cloetens1996phase, paganin2002simultaneous}, grating-based interferometry \cite{weitkamp2005x}, crystal analyzer-based methods \cite{kitchen2008simultaneous}, ptychography \cite{pfeiffer2018x}, single-shot grid imaging \cite{bennett2010grating, morgan2011quantitative}, and speckle-based imaging \cite{zdora2018state}, the latter also more generally referred to as modulation-based imaging \cite{quenot2022x, savatovic2025high}.

While all these methods significantly enhance tissue contrast and reveal those structural details that are difficult to access with X-ray attenuation imaging alone, they are currently unable to provide the tissue specificity that conventional histopathology delivers. In optical microscopy, pathologists rely on a wide range of dyes that selectively highlight different cellular components such as nuclei, cytoplasm, or specific proteins. To bridge this gap, X-ray-compatible stains that target specific tissues and cellular structures have been developed \cite{silva2015three, martins2017dual, busse2018three, muller2018nucleus, taphorn2022investigation}. Notable examples include modified versions of hematein and eosin \cite{muller2018nucleus, busse2018three}, the two primary stains used in conventional pathology to highlight cell nuclei and cytoplasm respectively. These X-ray stains work by selectively altering the X-ray interaction properties of targeted tissue components.

Since the recorded X-ray images always contain a combination of contrast or signal generated by the stain and the tissue, the precise location and quantity of the stain in each voxel are not directly accessible. Recently, it has been shown that this challenge can be addressed through material decomposition, a technique that separates different materials within a sample based on their distinct X-ray interaction properties. In a study imaging the inside of a cell with high resolution \cite{taphorn2022x}, the stain concentration was successfully retrieved by exploiting the unique material signatures of iodine and bromine.

Material decomposition traditionally relies on measurements at multiple X-ray energies \cite{alvarez1976energy}, though the same principle can be applied using other complementary measurements: X-ray dark-field signals at different energies \cite{sellerer2020dual}, or combined attenuation and phase-contrast measurements at a single energy \cite{kitchen2011phase, braig2018direct, sellerer2020dual, groenendijk2020material, schaff2020material, taphorn2022x}. This latter approach separates materials based on their unique combination of X-ray attenuation coefficients and electron number density.

In this work, we propose a novel method for obtaining three-dimensional contrast agent distributions in millimeter-sized biological tissues alongside morphological details on the micrometer scale. We achieve this using modulation-based X-ray imaging, which is based on analyzing the sample-induced distortions of a reference pattern imprinted onto the beam. While random masks like sandpaper \cite{morgan2012x, zdora2018state} or biological membranes \cite{berujon2012two} can be used to create a reference pattern, we use structured phase masks \cite{rizzi2013x, morgan2013sensitive, gustschin2021high} that increase measurement efficiency due to the stronger and more regular modulations produced \cite{gustschin2021high}. Different algorithms can be employed to track the sample-induced pattern distortions locally \cite{berujon2016x, zdora2017x} or globally \cite{quenot2021implicit, alloo2023m} and provide complementary images related to different sample properties (e.g. attenuation, small-angle scattering, refraction). Algorithms that simultaneously retrieve the sample's X-ray attenuation and electron number density---e.g. unified modulated pattern analysis (UMPA) \cite{zdora2017x}---are well-suited for material decomposition. However, phase effects originating at borders between adjacent materials are not accounted for in existing analyses, and strongly compromise quantitative accuracy given high spatial coherence. Though correction methods for these phase artifacts have been proposed \cite{wang2017speckle, groenendijk2020material, giakoumakis2022artifacts}, a successful correction with demonstrated quantitative results has not been reported until now.

This paper first presents tomographic X-ray measurements using materials with known X-ray properties, confirming that modulation-based imaging can be used to obtain quantitatively accurate phase and attenuation measurements and perform material decomposition. In a second experiment, we examine a mouse kidney stained with a hematein-lead complex \cite{muller2018nucleus}, which is a promising stain that creates contrast in both in the visible light regime (due to the dye properties of hematein) and the X-ray regime (due to the high X-ray attenuation of lead). We show that after application of a novel edge-artifact correction, the complementary electron number density and X-ray attenuation volumes may be used to separate the signals stemming from the contrast agent and the soft tissue. This provides access to the three-dimensional lead concentration within the sample alongside high-contrast morphology. We then validate our quantitative results against K-edge subtraction imaging, a standard method for element-specific concentration measurement.
In a third experiment, we perform a modulation-based tomographic scan on a sample stained with a lower, histologically compatible concentration of the stain. We use the dual optical-X-ray properties of the contrast agent to compare the retrieved stain concentrations to a slice through the sample obtained with conventional, optical histology. In this manner, we create a direct bridge between virtual and conventional histopathology.

\section{Results and Discussion}
\subsection{Quantitative material decomposition}
\begin{figure}[h!]
    \centering    
    \includegraphics[width=1\linewidth]{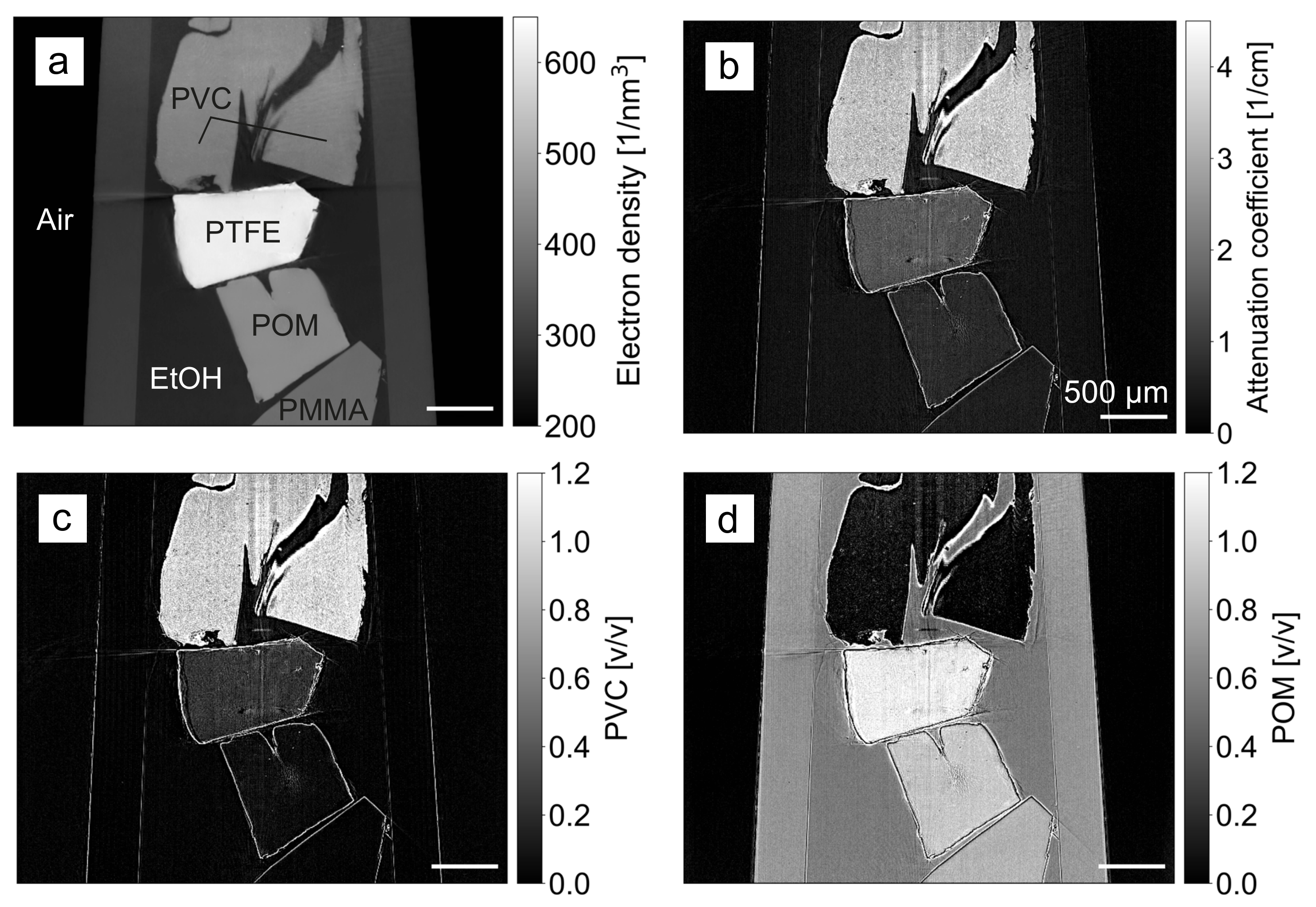}
    \caption{Central slice through tomographic reconstructions of the test sample. (a) shows the electron number density, (b) the linear attenuation coefficient. (c) and (d) show a material decomposition into PVC and POM volume fractions. As expected, the volume fraction of the target materials is close to 1 within the corresponding part of the sample.}
    \label{fig:material-decomposition-phantom}
\end{figure}

To confirm the quantitative accuracy of modulation-based imaging, a tomographic scan was performed on a test sample consisting of materials with known properties. To span a range of material properties, including some similar to soft tissue, the materials chosen were polyoxymethylene (POM), polyvinyl chloride (PVC), poly(methyl methacrylate) (PMMA) and polytetrafluoroethylene (PTFE), all submerged in pure ethanol. A slice through the sample is shown in Fig.~\ref{fig:material-decomposition-phantom}; the differences in material properties are clearly discernible in both electron number density (a) and X-ray attenuation coefficients (b).

The averaged attenuation and electron number density values for each material were obtained as described in Sec.~\ref{sec:calibration} and are shown in Fig.~\ref{fig:material-decomposition-phantom}. These values are in good agreement with the theoretical values, demonstrating that the method accurately retrieves the physical quantities. This finding is consistent with previous work using a random diffuser as a modulator \cite{zandarco2024speckle}.
Values at the edges of the materials were excluded from the analysis due to their amplified signal, a matter that will be addressed in Sec.~\ref{sec:virtual-histology}.

The two complementary signals enable material decomposition by defining a new coordinate system based on the theoretical electron number density and attenuation coefficient values for POM and PVC, which were chosen because the properties of all the materials in the sample fall within the span of this system. As shown in Fig.~\ref{fig:material-decomposition-phantom}(c) and (d), the volume fraction of POM is close to 1 in the POM image and the fraction of PVC is close to 1 in the PVC image, which is the desired behavior. The other materials appear in different shades depending on their similarity to the base materials. Note that in this example, the volume fraction of PTFE exceeds 1. This is a mathematical consequence of its high electron number density, which causes it to be represented as an excess fraction relative to the base materials.

\begin{figure}[h!]
    \centering
    \includegraphics[width=1\linewidth]{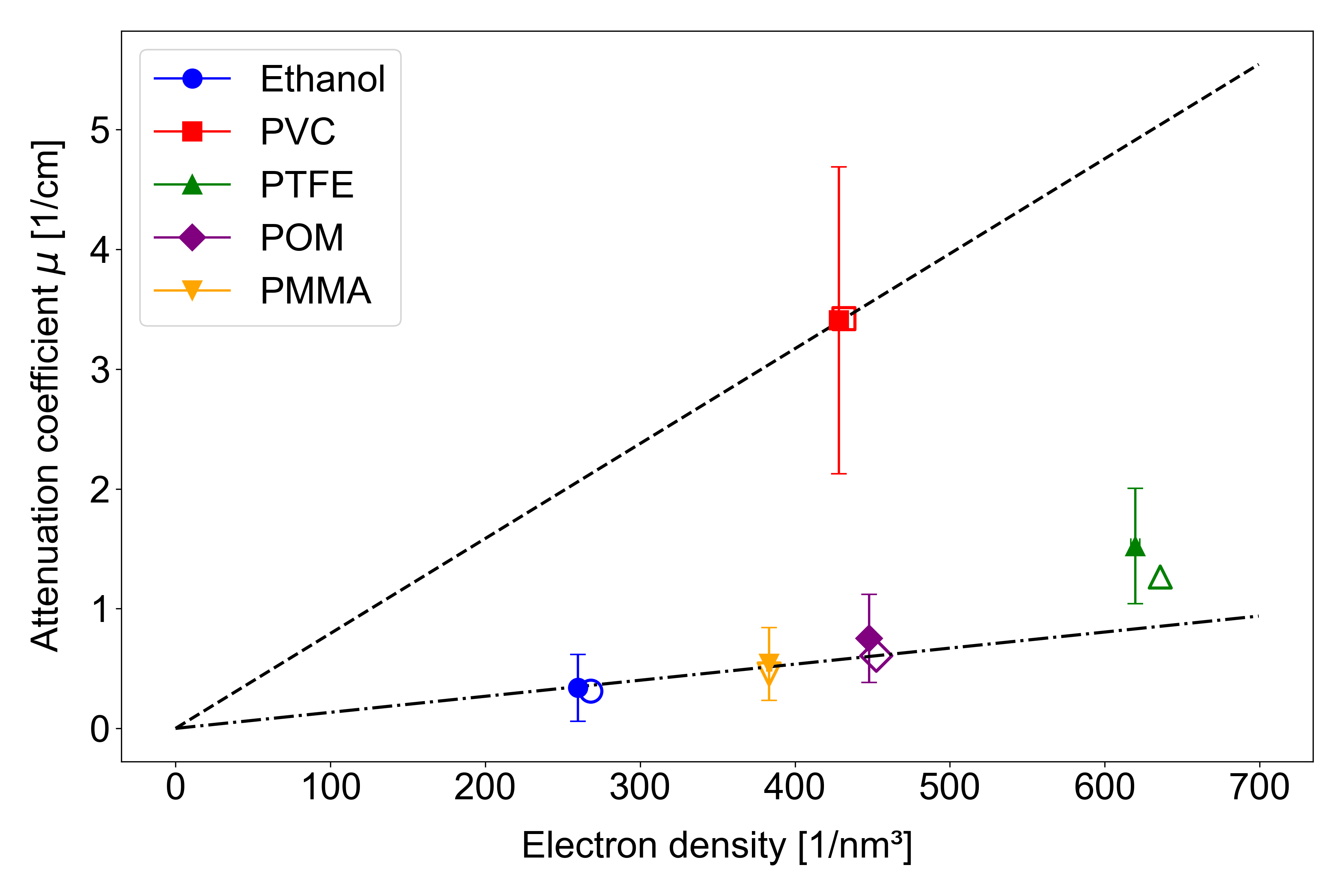}
    \caption{Comparison of theoretical and measured X-ray attenuation coefficients and electron densities for different materials in the sample shown in Fig.~\ref{fig:material-decomposition-phantom}, shown with open and filled data point symbols respectively. The horizontal uncertainties are in the range of the measurement points. The dotted line and dash-dotted line indicate the coordinate systems spanned by the theoretical values of PVC and POM, respectively, used for material decomposition. Note that these lines serve as reference axes only, and are not linear fits; the proximity of ethanol and POM to one of the axes is coincidental.}
    \label{fig:calibration}
\end{figure}

\subsection{Quantitative stain mapping}
\label{sec:virtual-histology}
\begin{figure}
    \centering
    \includegraphics[width=1.0\linewidth]{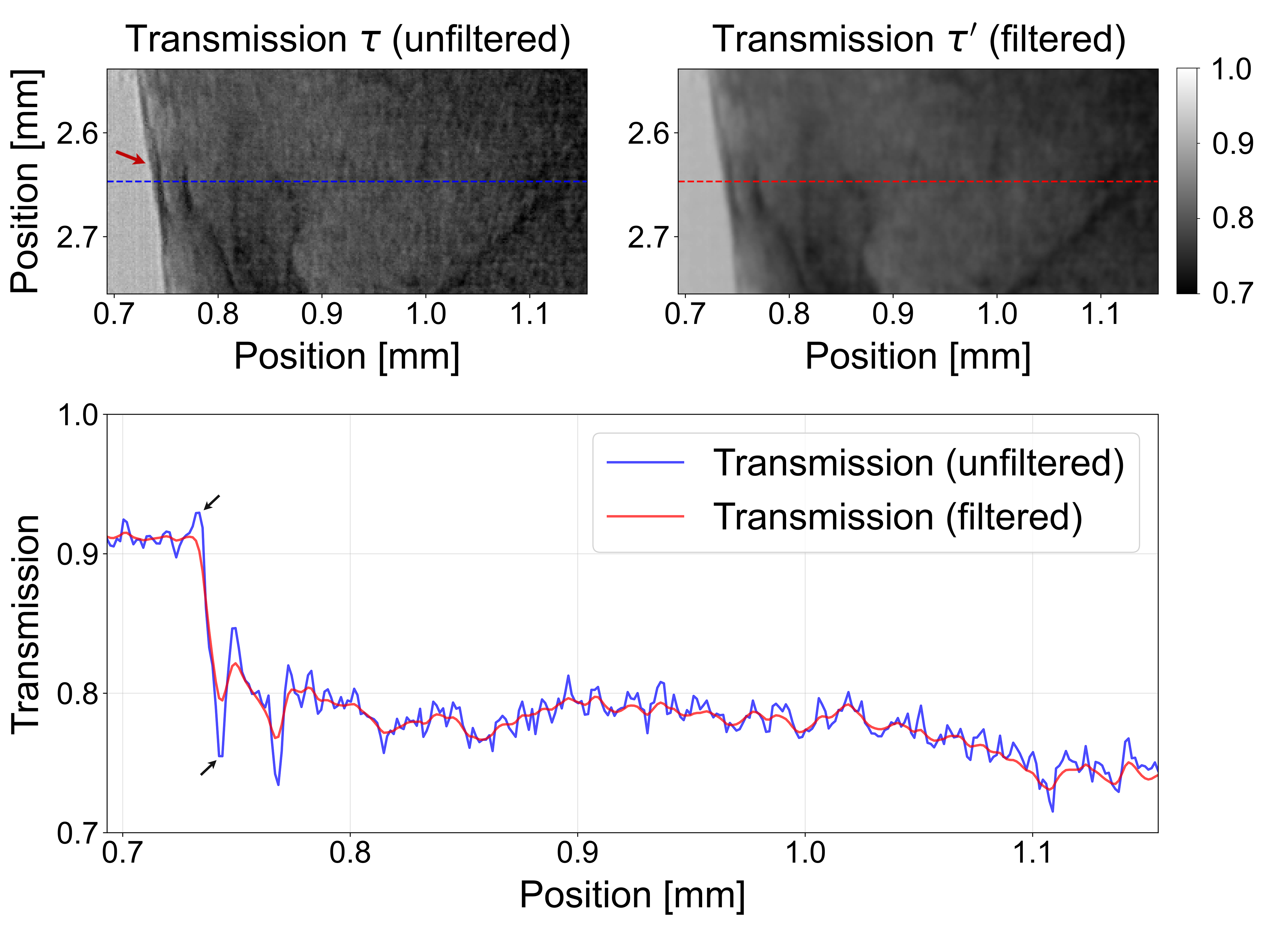}
    \caption{Comparison of cropped regions of the unfiltered transmission image $\tau$ (top left) and filtered transmission image $\tau'$ (top right) of the mouse kidney sample. The line plot (bottom) corresponds to the values across the blue and red lines in the images, averaged over two adjacent rows of pixels. A reduction in Laplacian phase effects is observed with the application of the filter, most notably in the transition from the ethanol to the stained tissue (black arrows).}
    \label{fig:line-plot}
\end{figure}

To demonstrate the capabilities of our proposed virtual histology approach, a tomography scan was performed on a mouse kidney stained with a hematein-lead complex. Since the attenuation signal would otherwise show amplified signal at tissue borders due to phase effects, as described in the previous section, all transmission projections were first corrected following the method described in Sec.~\ref{sec:edge-correction}. For this purpose, an unstained section of the same kidney was scanned under identical conditions and used to estimate the electron number density and linear X-ray attenuation coefficient of native tissue. These values and the X-ray interaction properties of lead provided the required parameters for a physics-informed correction \cite{paganin2002simultaneous}. As shown in Fig.~\ref{fig:line-plot} on a zoomed-in transmission projection, this correction removes the amplified contrast at borders while improving the noise characteristics of the image (cf.~Ref.~\cite{gureyev2017unreasonable}). The difference is most visible at the ethanol-tissue border (red arrow), which shows the expected direct transition from ethanol to tissue grey values only after filtering (black arrows indicating uncorrected edge contrast).

\begin{figure*}
    \centering
    \includegraphics[width=1.0\linewidth]{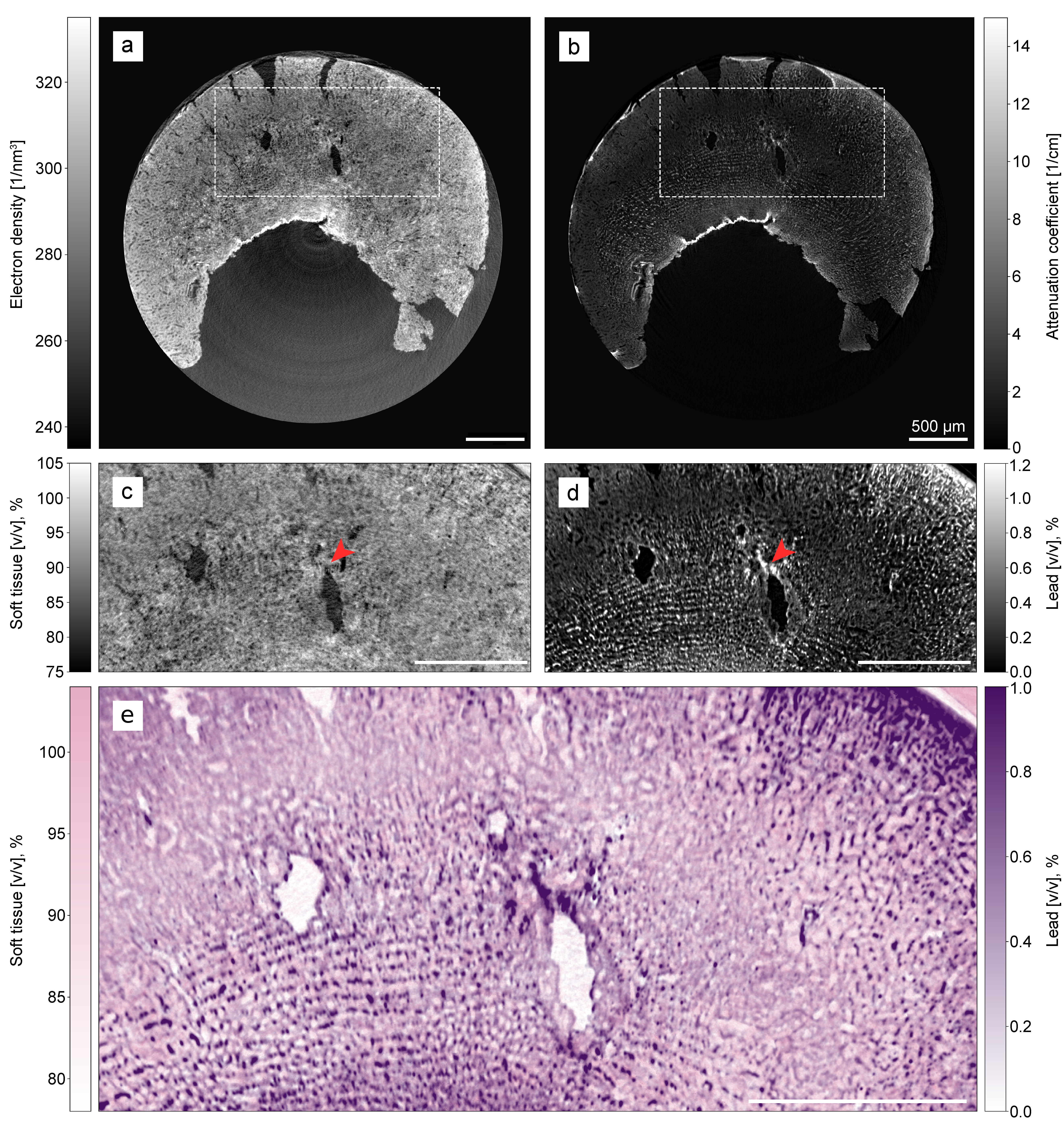}  
    \caption{Slices through the tomographic reconstruction of the lead-stained mouse kidney sample. (a) Electron number density. (b) Linear X-ray attenuation coefficients (after removal of edge-enhancement effects). (c) Soft tissue-only image calculated based on data from the two regions of interest in the first row. (d) Lead volume fractions calculated analogously. (e) Merged soft tissue and lead volume fraction image. False colors are applied to mimic conventional histology images, where most cell structures appear pink due to eosin while cell nuclei are enhanced in purple. Red arrows indicate an example region showing a high electron number density in (a) mainly due to the presence of stain, which appears with reduced contrast in the corresponding soft tissue volume fraction image (c) since the staining contribution has been separated out into the lead volume fraction image (d). The sample container walls were removed using a mask in (a) and (b).}
    \label{fig:lead-rat-kidney}
\end{figure*}

Tomographic reconstruction was then performed using the corrected attenuation projections and the original phase projections. Slices through the resulting electron number density and attenuation coefficient volumes are shown in Fig.~\ref{fig:lead-rat-kidney}(a) and \ref{fig:lead-rat-kidney}(b). Based on these signals, we apply material decomposition (see Sec.~\ref{sec:material-decomp}) to retrieve a soft-tissue volume (c) and a lead volume (d), defined by the X-ray interaction properties of unstained kidney and lead, respectively. This results in a three-dimensional map of the lead volume fraction, creating access to the concentration of the stain in each voxel. The material decomposition also provides a soft tissue-only image, which shows the morphology of the sample without the effects of the contrast agent. The complementarity of the two images is most visible in the area marked by the red arrows in Fig.~\ref{fig:lead-rat-kidney}, which shows a strong signal in the lead image but not in the soft tissue image. We then use a false-color scheme inspired by conventional histology to overlay the lead concentrations (purple) on top of the soft tissue image (pink). Using this scheme, the latter signal provides a backdrop of contrast by highlighting the general morphology, mimicking the role of the eosin stain in conventional histology. The superimposed lead map corresponds to the position of hematein, which is colored in violet. The resulting image is displayed in Fig.~\ref{fig:lead-rat-kidney}(e) and demonstrates the selective highlighting by the stain, primarily taken up by cell nuclei. It is likely, however, that the high lead concentrations at the top right border of the tissue are not due to a strong presence of cell nuclei, but rather due to excess stain that was not removed in the washing step.

\begin{figure}
    \centering
    \includegraphics[width=1.0\linewidth]{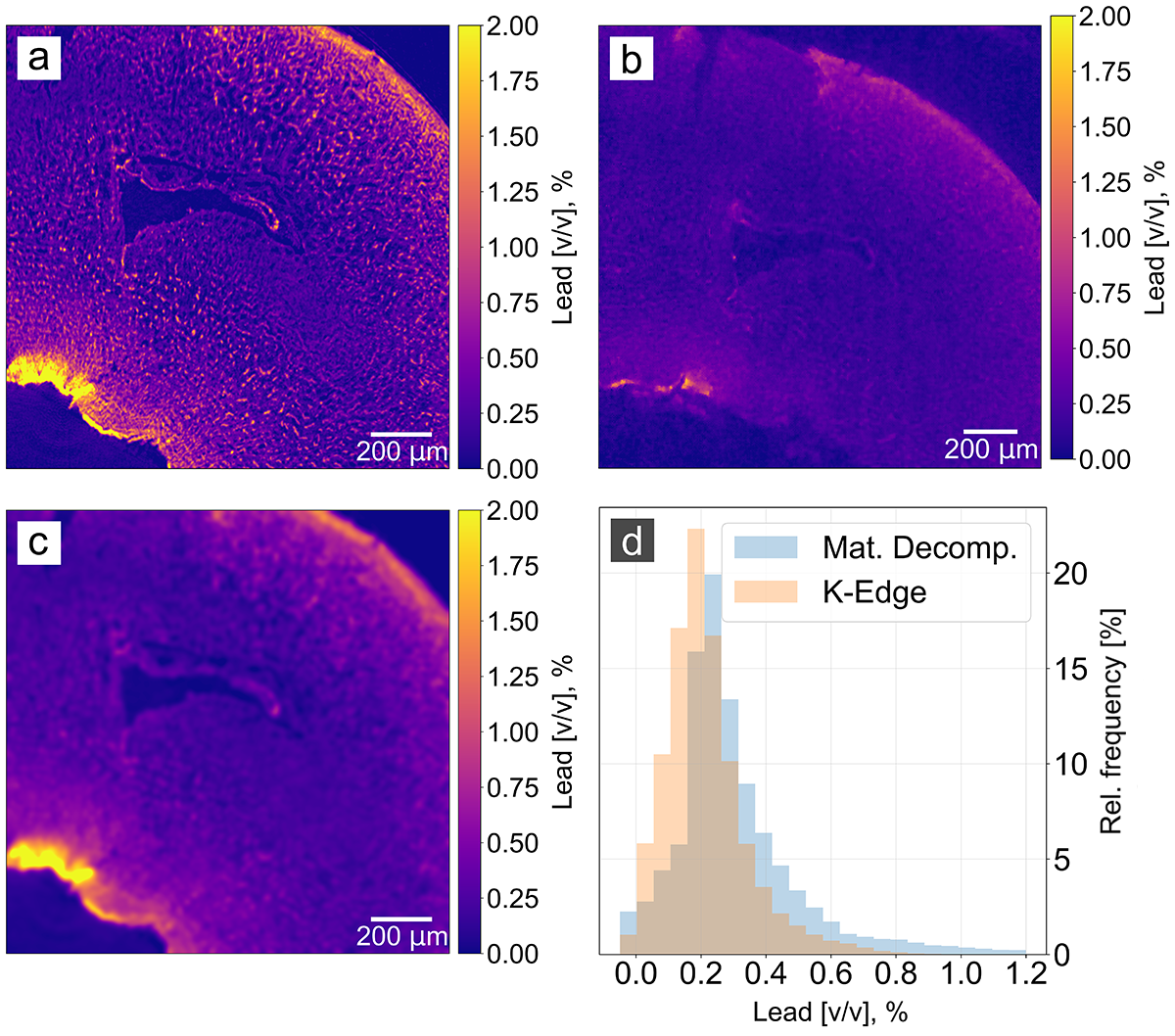}
    \caption{Comparison of lead concentrations obtained via material decomposition and K-edge subtraction imaging. (a) shows the lead concentration according to the material decomposition, (b) the K-edge subtraction image of a similar slice in the volume. (c) shows a blurred version of (a) to account for the lower spatial resolution of (b). (d) shows histograms of (b) and (c) in comparison and demonstrates general agreement of the methods with slightly lower values for the K-edge method, possibly due to stain leaching between scans.}
    \label{fig:comparison-of-scans}
\end{figure}

To confirm the quantitative accuracy of the lead concentrations retrieved using our approach, we compare the presented results to those acquired using the established method of K-edge subtraction imaging (see Sec.~\ref{sec:k-edge-subtraction}). Figure \ref{fig:comparison-of-scans} shows slices from the three-dimensional concentration maps obtained using (a) the material decomposition approach and (b) the K-edge approach. The slices are taken from similar locations in the kidney. Compared to the material decomposition approach in (a), the concentration map in (b) exhibits higher noise and lower spatial resolution. This is due to the lower sample contrast and decreased beamline flux at the high energies on the order of \SI{90}{\kilo\eV} required for this measurement. However, application of a Gaussian blur (kernel size 41) to (a) to match the spatial resolution of the K-edge image creates the image in panel (c), showing that the two methods are generally in good agreement. This visual impression is also confirmed by the histogram in panel (d). The main discrepancy occurs at tissue borders, where K-edge imaging detects less lead. This difference may be due to sample degradation since parts of the ethanol-preserved specimen became brittle and detached during transport between the two measurements. Additionally, the fresh ethanol used for the K-edge measurement may have leached some stain.

Compared to K-edge measurements, our material decomposition approach has the advantage of not requiring an energy switch during data acquisition. Another advantage is that morphology is retrieved alongside concentrations, which is difficult when soft tissue contrast is low at the K-edge energy. Furthermore, the phase and attenuation volumes obtained through our method are precisely spatially correlated since they originate from the same measurement, thus eliminating registration errors. In contrast, the two volumes measured above and below the K-edge energy in conventional approaches may require spatial registration before analysis if sample drift or movement occurs, which can compromise quantitative accuracy.

\subsection{Comparison to conventional histology}
\begin{figure*}
    \includegraphics[width=1.0\linewidth]{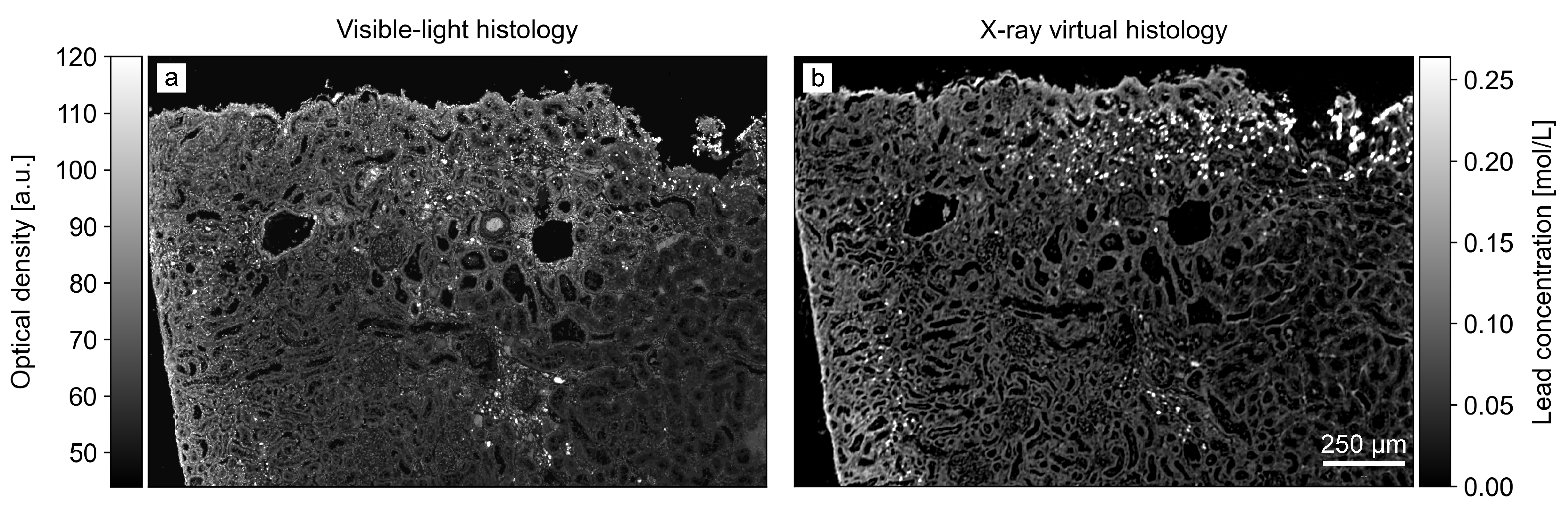}
    \caption{Visual comparison of (a) visible-light microscopy image, converted to greyscale, with (b) a slice through the three-dimensional concentration volume in a similar location. The similarity is explained by the dual optical-X-ray properties of the stain, allowing it to create contrast both in the visible light and X-ray regime. While (a) allows distinguishing finer features better, (b) provides quantitative values for the stain concentration.}
    \label{fig:virtual-vs-real-histology}
\end{figure*}

\label{sec:comparison-histology}
In a third experiment, a rat kidney was treated with a lower stain concentration than in Sec.~\ref{sec:virtual-histology} and embedded in a block of paraffin wax. The stain concentration was chosen closer to the values used in conventional histology protocols to enable subsequent analysis of the specimen under an optical microscope. A computed tomography scan was performed on the rat kidney sample; using material decomposition, a three-dimensional volume of the stain concentration was then calculated. After the scan, a histological section of the sample was obtained and converted into a greyscale optical density image.

The examined X-ray stain is a coordination complex that combines hematein, an optical dye, with lead ions. For this reason, the presence of the stain creates contrast both in the visible light regime and the X-ray regime. This property becomes apparent in Fig.~\ref{fig:virtual-vs-real-histology}, where the visible-light histology image is compared to a similar slice in the three-dimensional concentration volume. The slice through the volume was obtained through the 2D-3D registration algorithm (see Sec.~\ref{sec:comparison-optical-histology}); note that a perfect registration was not possible due to deformations introduced by the physical cutting process.

The result shows very good visual agreement between the slice obtained using optical histology and the virtual concentration map, demonstrating the success of the described material decomposition approach. While finer features are better resolved in the visible-light microscopy image, the X-ray virtual histology approach enables non-destructive retrieval of stain concentrations in three dimensions. Using Fourier ring correlation (see Sec.~\ref{sec:comparison-optical-histology}), the spatial resolution in the axial plane of the X-ray virtual histology dataset was determined to be \SI{5.67(0.02)}{\micro\metre} (half-bit criterion) or \SI{6.42(0.01)}{\micro\metre} (full-bit criterion), which is on the order of the size of mammalian nuclei of 5--$\SI{10}{\micro\metre}$ \cite{li2011central}.

\section{Conclusion \& Outlook}
In this work, we have developed an imaging approach that enables quantitative and therefore comparable virtual histology of stained tissue, with resolution in the micrometer range. Our X-ray modulation-based method allows precise measurement of contrast agent concentrations alongside tissue morphology, providing objective metrics for tissue assessment that could reveal variations in agent distribution potentially linked to pathological changes.
Furthermore, we have validated our approach against conventional histology using the properties of an X-ray stain that creates contrast in both the visible-light and X-ray regimes. In this manner, we have created a direct bridge from volumetric X-ray data to conventional two-dimensional histology. This approach may be useful for the development of X-ray stains and the three-dimensional analysis of stained tissue samples without cutting.

Beyond standalone applications, our virtual histology approach complements conventional histology methods: The three-dimensional X-ray data provides a complete overview of the tissue sample, allowing researchers to identify regions of interest before physical sectioning. This guided approach can improve the efficiency of conventional histological analysis by ensuring that the most relevant areas are selected for high-resolution optical microscopy. Although our approach shows promise for tissue characterization, achieving the same spatial resolution as conventional histology remains challenging for X-ray microscopy. While the resolution achieved in Sec.~\ref{sec:comparison-optical-histology} is on the order of the size of mammalian cell nuclei, this does not yet allow for a detailed inspection of individual nuclei, as is possible in conventional histology. While increased measurement time can be used to improve spatial resolution, it is ultimately limited by the point spread function (PSF) of the detector. We note here that initial work has been done to surpass this limit through a super-resolution approach \cite{gunther2019full}.

Looking forward, there are several avenues for further research that could extend this work. Our approach may be adopted for use with compact inverse-Compton-based sources \cite{zandarco2024speckle}, liquid-metal-jet sources \cite{zanette2014speckle}, or conventional X-ray sources \cite{quenot2021implicit}. Additionally, alternative contrast agents should be explored, and incorporating the x-ray dark-field signal could allow for the use of nanoparticle-based contrast agents in addition to conventional stains \cite{navarrete2024nanoparticle}. The ability to generate large datasets of virtual histology images may also open opportunities for automated analysis using AI models, enabling high-throughput screening and quantitative assessment of tissue features. 

\section{Materials \& Methods}
\subsection{Sample preparation}
\label{sec:stain}
For the experiments in Sec.~\ref{sec:virtual-histology}, a wildtype mouse kidney was extracted, fixed in formalin, and cut into six pieces, two of which were used for this study. One piece was stained according to the hematein-lead X-ray stain protocol described in Ref.~\cite{muller2018nucleus}, while another piece was left in its native state. Hematein staining is a standard procedure in conventional histology, used to highlight cell nuclei in blue. The stain applied in this study is a modified version that uses lead acetate (rather than a lighter metal) as the intermediate binding partner between the negatively charged DNA backbone and hematein. This creates additional X-ray contrast while retaining the binding properties of the original stain \cite{muller2018nucleus}. Immediately before the scanning, the samples underwent a dehydration process starting with a concentration of \SI{70}{\percent}~vol.~ethanol. The samples were finally scanned inside a sealed pipette tip containing \SI{100}{\percent} ethanol to prevent radiolysis-induced bubbles that appear in aqueous solvents. A piece of PMMA was included in the pipette tip and acted as a calibration material for the phase measurements due to its known electron number density (see Sec.~\ref{sec:calibration}). The stained sample in Figs. \ref{fig:line-plot}-\ref{fig:comparison-of-scans} was analyzed using K-edge subtraction imaging at the Deutsches Elektronensynchrotron (DESY), Germany, approximately six months after the modulation-based imaging at the Australian Synchrotron. Note the sample was stored in the same solution for these 6 months, and then transferred into a new sample container filled with \SI{100}{\percent} ethanol. 

The samples for the above experiments came from a female C57BL/6J mouse. Housing and dissection of organs was carried out at the neurophysiology section of the biological department, University of Hamburg, and in accordance with European Union’s and local welfare guidelines (Behörde für Gesundheit
und Verbraucherschutz, Hamburg, Germany; GZ G21305/591-00.33).

For the experiments in Sec.~\ref{sec:comparison-histology}, a piece of a rat kidney sample was prepared with a different protocol to achieve a more histologically compatible staining with lower concentrations, inspired by the method described by Metscher \cite{metscher2021simple}. Two solutions were prepared: solution A consisted of \SI{66}{\milli\molar} lead(II) acetate trihydrate in distilled water (approximately \SI{2.5}{\gram} Pb(II) acetate in \SI{10}{\milli\liter} $\mathrm{dH_2O}$), while Solution B contained \SI{1}{\percent} hematein in \SI{20}{\percent} ethanol (\SI{1}{\gram} hematoxylin, \SI{0.2}{\gram} sodium iodate, \SI{20}{\milli\liter} absolute ethanol, and \SI{80}{\milli\liter} $\mathrm{dH_2O}$, oxidized at room temperature for \SI{24}{\hour} without lid). Prior to staining, the sample was washed for \SI{24}{\hour} in distilled water to reduce precipitates according to the recommendations in Ref.~\cite{metscher2021simple}. Solutions A and B were then mixed in equal parts (\SI{1}{\milli\liter} + \SI{1}{\milli\liter}), and the sample was transferred from the washing solution to the staining solution, where it remained for approximately \SI{24}{\hour} on a shaking plate. During this period, the sample was briefly removed for a laboratory-based preview CT scan before being returned to the staining solution. Following staining, the sample was washed again in $\mathrm{dH_2O}$ to remove excess stain and stored for more than \SI{24}{\hour} until further processing. Finally, the sample underwent dehydration and embedding in paraffin for subsequent analysis. It was first scanned at the P05 beamline of DESY and subsequently analyzed histologically. For this purpose, multiple slices of thickness \SI{5}{\micro\metre} were extracted using a microtome. The paraffin wax was removed using xylene, followed by incubation in a descending ethanol series. The slices were then briefly subjected to tap water and subsequently dehydrated using an ascending ethanol series back to xylene. Finally, the sections were mounted on glass slides with a coverslip.

The animal housing and organ removal for this sample was carried out at Helmholtz-Zentrum Munich following the European Union guidelines 2010/63. The procedure was conducted in compliance with the ethical standards of the institution and approved by the responsible governmental body. Specifically, the experiments were performed under the license number ROB-55.2-2532.Vet\_02-21-133.



\subsection{Experimental setup}
\begin{figure}
    \centering
    \includegraphics[width=1.0\linewidth]{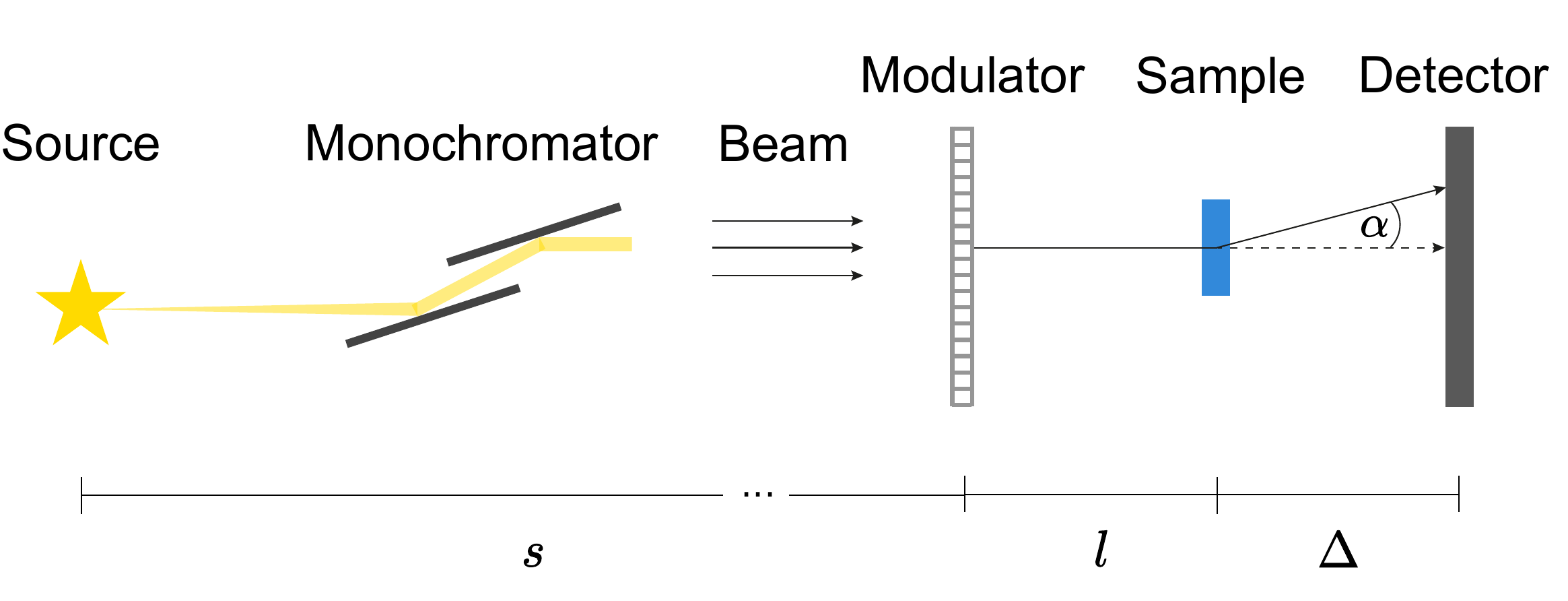}
    \caption{The beam originates from a bending magnet source (MCT) or an undulator (P05) and is set to the desired energy using a double multi-layer monochromator. After a distance $s$, the beam impinges upon the modulator, which introduces a phase shift in regular intervals across the wavefront. The sample is positioned at a distance $l$ from the modulator, and the detector is at a distance $\Delta$ from the sample.}
    \label{fig:setup-sketch}
\end{figure}
The material decomposition experiment described in Sec.~\ref{sec:virtual-histology} was performed at the microCT beamline (MCT) of the  Australian Synchrotron, a third-generation light source operating with an electron energy of \SI{3}{\giga\eV} and a beam current of \SI{200}{\milli\ampere} in the storage ring \cite{arhatari2023micro}. A sketch of the setup is shown in Fig.~\ref{fig:setup-sketch}: the radiation originates from a \SI{1.3}{\tesla} bending magnet and is then filtered to an energy of \SI{25}{\kilo\eV} using a double-multilayer monochromator. After a distance $s = \SI{21.3}{\metre}$, the beam impinges on a Talbot array illuminator \cite{gustschin2021high} made of silicon with a period of \SI{10}{\micro\metre} and a design height of \SI{25.7}{\micro\metre}, which we refer to as the modulator. This height corresponds to introducing a $2\pi/3$ phase shift at the design energy of \SI{30}{\keV}. The sample was placed $l = \SI{0.19}{\metre}$ downstream of the modulator and was followed by the sCMOS detector placed a further \SI{0.16}{\metre} downstream. For the camera, a $4.5 \times$ magnification lens was used, leading to an effective pixel size of \SI{1.44}{\micro\metre} at the detector position and a field of view of 3.7 $\times$ \SI{3.1}{\milli\metre\squared} (width $\times$ height).

For each scan, 9 consecutive tomographic acquisitions were performed with an exposure time of \SI{40}{\milli\second} per projection. For each tomography, the modulator was placed in a different transverse position. The signal retrieval was performed using the local feature-tracking algorithm UMPA \cite{zdora2017x}, specifically a runtime-optimized version \cite{de2022high} with a window size parameter of 1 (effective window size $3 \times 3$ pixels). The electron number density and attenuation volumes were retrieved from the UMPA-processed projections with the FBP algorithm of the \emph{Core Imaging Library (CIL)} \cite{jorgensen2021core}, which uses the \emph{ASTRA Toolbox}  \cite{van2016fast} as its backend.

The calibration sample was scanned using 1800 projections acquired over an angular range of $180^\circ$. Because the biological samples were too large to fit the detector's horizontal field of view, they were scanned with an offset rotation axis in 3600 projections over an angular range of 360$^\circ$ (see Sec.~\ref{sec:offset-rotation-axis}).

Additionally, the lead-stained mouse kidney sample was measured at the High Energy Materials Science Beamline P07, which is operated by Helmholtz-Zentrum Hereon at PETRA III, DESY in Hamburg, Germany \cite{schell2014high}. The measurement was conducted at this beamline because the K-edge energy of lead (\SI{88}{\kilo\eV}) falls outside the operating specifications of the MCT beamline. The P07 beamline enabled two consecutive absorption scans at a propagation distance of \SI{3}{\centi\metre} and energies of \SI{87}{\kilo\eV} and \SI{89}{\kilo\eV}, respectively. The detector has an array of $7920 \times  6004$ pixels with a pitch of \SI{4.6}{\micro\metre}, which leads to an effective pixel size of \SI{0.90}{\micro\metre} using $5.1$-fold magnification. 5000 projections were acquired over 180$^\circ$ for each scan with an exposure time of \SI{350}{\milli\second} each. The sample fit into a single field-of-view, meaning no stitching was required. To correct for a slight deformation of the sample during the scan, the volumes captured at two energies were deformably registered with the \emph{ImFusion} software before quantitative analysis \cite{imfusion2024}. Additionally, both volumes were bilaterally filtered individually to reduce the impact of noise while retaining edge sharpness. Before volumetric reconstruction analogously to the previous experiment, the data was binned 5-fold to reduce the impact of noise. 

The material decomposition experiment described in Sec.~\ref{sec:comparison-histology} was conducted at the Imaging Beamline P05, which is operated by Helmholtz-Zentrum Hereon at PETRA III, DESY in Hamburg, Germany \cite{wilde2016micro}. An undulator source was used in combination with a double crystal monochromator to create a highly coherent and monochromatic beam with an energy of \SI{20}{\kilo\eV}. The detector uses a CMOSIS CMV 20000 sensor with $5120 \times 3840$ pixels and a pitch of \SI{6.5}{\micro\metre}, leading to an effective pixel of \SI{1.28}{\micro\metre} with the 5-fold magnification objective. The experimental setup is similar to the setup at the microCT beamline used for Sec.~\ref{sec:virtual-histology}. The modulator was a Talbot array illuminator with period of \SI{7}{\micro\metre} and a design height that introduces a $2\pi/3$ phase shift at its design energy of \SI{20}{\keV}. Both the distances $l$ and $\Delta$ were equal to $\SI{0.175}{\metre}$. 3001 projections with an exposure time of \SI{80}{\milli\second} each were acquired over an angular range of \SI{180}{\degree} for each of the 16 modulator positions. A dynamic reference correction was performed by decomposing the recorded reference images into eigencomponents \cite{van2015dynamic}. Each projection was then corrected with the linear combination of eigencomponents that yielded the best least-squares fit in a sample-free region, as described in Ref.~\cite{riedel2023comparing}. The signal retrieval was then performed using UMPA \cite{zdora2017x} with a window size parameter of 1 (effective window size $3 \times 3$ pixels). The tomographic reconstruction was performed using the FDK algorithm implemented in the \emph{Xaid} software suite (\emph{Mitos GmbH}, Garching b. München, Germany).

\subsection{Accurate retrieval of X-ray attenuation in two-material mixtures}
\label{sec:edge-correction}
In conventional X-ray imaging, the two-dimensional transmission image $\tau(x,y)$ represents the logarithmic ratio of transmitted intensity $I(x,y)$ to incident X-ray intensity $I_0(x,y)$, where $(x,y)$ are the real-space coordinates in the image plane. According to the Beer-Lambert law, the transmission is directly related to the sample's linear X-ray attenuation $\mu(x,y,z)$ coefficient via
\begin{equation}
\label{eq:simple-transmission}
    \tau(x,y) = -\ln\left(\frac{I(x,y)}{I_0(x,y)}\right) = \int_0^{T(x,y)} \mu(x,y,z) \, \text{d}z,
\end{equation}
where $T(x,y)$ is the thickness of the sample and $z$ refers to the coordinate along the beam propagation direction. When imaging samples with coherent X-ray sources and a non-negligible sample-to-detector distance, however, the image retrieved by application of Eq.~\eqref{eq:simple-transmission} contains additional contributions related to the Laplacian of the sample-induced x-ray wavefield phase shift $\Delta \phi(x,y)$. These manifest as increased contrast at the material boundaries. While this increased contrast benefits methods such as propagation-based imaging \cite{paganin2002simultaneous, gureyev2017unreasonable} or implicit speckle-tracking  \cite{alloo2023recovering}, it is problematic for accurate material decomposition if left unaccounted for, since material decomposition assumes a clear separation between the phase and attenuation signal (see Sec.~\ref{sec:material-decomp}).
Although such Laplacian phase effects can be simulated and removed \cite{groenendijk2020material}, this method requires manual tuning of parameters, and any mismatch can introduce noise. In this section, we therefore propose an alternative approach that takes into account the Laplacian phase effects to obtain an image solely related to the sample's attenuating properties.

The Paganin filter, as first proposed by Paganin et al. \cite{paganin2002simultaneous}, is a method for optics-free phase-retrieval using a single propagation distance and the assumption that the sample consists of a single material. The strength of the filter depends on the $\delta/\mu$-ratio of the material, where $\delta$ is its refractive index decrement and $\mu$ is its linear attenuation coefficient.
It has been shown that the method can be extended to multiple-material samples by considering a piece of material 1 embedded in a different material 2 \cite{gureyev2002quantitative}. The $\delta/\mu$-ratio is then replaced by a relative ratio
\begin{equation}
\label{eq:relative-delta-beta}
    \left(\frac{\delta}{\mu}\right)_\text{rel} = \frac{\delta_1 - \delta_2}{\mu_1 - \mu_2},
\end{equation}
where $\delta_i$ and $\mu_i$ refer to the refractive index decrement and linear attenuation coefficient of each material \cite{gureyev2002quantitative, beltran20102d}. We extend this approach to arbitrary mixtures of materials by first assuming that each voxel of the reconstructed volume consists of a mixture of two materials, a fraction $f$ of material 1 and a fraction $(1-f)$ of material 2. The effective $\delta$ and $\mu$ of a given voxel are then the linear combinations
\begin{equation}
    \delta_\text{mix} = f\,\delta_\text{1} + (1-f)\, \delta_\text{2}
\end{equation}
and
\begin{equation}
    \mu_\text{mix} = f\,\mu_\text{1} + (1-f)\, \mu_\text{2}
\end{equation}
(cf.~Ref. \cite{gureyev2015monomorphous}). Assuming such a mixture with a fraction $f_1$ is embedded within a second mixture of fraction $f_2$, the relative $\delta/\mu$ ratio using Eq.~\eqref{eq:relative-delta-beta} is
\begin{align}
\left(\frac{\delta}{\mu}\right)_\text{rel} &= \frac{[f_1\delta_\text{1} + (1-f_1)\delta_\text{2}] - [f_2\delta_\text{1} + (1-f_2)\delta_\text{2}]}{[f_1\mu_\text{1} + (1-f_1)\mu_\text{2}] - [f_2\mu_\text{1} + (1-f_2)\mu_\text{2}]} \nonumber \\
&= \frac{(f_1-f_2)(\delta_\text{1} - \delta_\text{2})}{(f_1-f_2)(\mu_\text{1} - \mu_\text{2})} = \frac{\delta_1 - \delta_2}{\mu_1 - \mu_2}.
\end{align}
The above equation shows that knowledge of the mixtures $f_1$ or $f_2$ is not required via this approach. Thus, the same Paganin filter is valid for arbitrary mixtures of two given materials.

For the case of single-distance propagation-based imaging, Croton et al.~have demonstrated how a pure attenuation image of a sample may be obtained by applying a suitably modified Paganin filter to the recorded intensity images \cite{croton2018situ}. This is achieved by assuming the object is made of two different materials, and the total projected thickness of the object varies slowly. This is valid for the samples examined in this work due to the cylindrical, ethanol-filled containers used as sample tubes.

We propose to extend the approach by Croton et al.~to modulation-based imaging, where a sample is imaged at multiple different modulator positions. An approximate uncorrected transmission image may be obtained by averaging over all sample images $I_i(x,y)$ and reference $I_{0, i}(x,y)$ images:
\begin{equation}
    \tau(x,y) = \sum_{i} \frac{I_i(x,y)}{I_{0, i}(x,y)}.
\end{equation}
This approximation assumes that by superimposing a sufficient number of different modulator positions, the effect of the mask on the image cancels out on average. The filter is then applied to this image (cf. Refs.~\cite{croton2018situ, pavlov2020single}) to obtain the corrected transmission image
\begin{equation}
    \label{eq:modified-filter}
   \tau'(x,y) = \mathcal{F}^{-1}\Bigg[\frac{\mathcal{F}\big[{\tau}(x,y)\big]}{1 + \big(\frac{\delta}{\mu}\big)_{\text{rel}}\,\Delta\,\mathbf{k}_{\perp}^2}\Bigg],
\end{equation}
where $\Delta$ is the distance between the sample and the detector. Here, $\mathcal{F}$ denotes the Fourier transformation with respect to $x$ and $y$ and $\mathcal{F}^{-1}$ is the corresponding inverse Fourier transformation; $\mathbf{k}_{\perp} = (k_x, k_y)$ denotes the Fourier-space coordinates corresponding to $(x,y)$. The resulting expression is similar to a previous result derived in the context of single-shot speckle-based imaging \cite{pavlov2020single}.

The resulting projection is thus corrected for propagation-based effects related to the Laplacian of the x-ray wavefield phase and accurately quantifies the transmission of X-rays through the sample; additionally, the amount of noise in the image is reduced \cite{gureyev2017unreasonable}. It shall be noted in this context that the filter term is derived based on the finite-difference approximation of the transport-of-intensity equation \cite{teague1983deterministic}, given by Refs.~\cite{paganin2002simultaneous, paganin2006coherent} as
\begin{equation}
I(x,y,z=\Delta) = \left(1 - \frac{\delta \Delta}{\mu} \nabla_\perp^2 \right) I(x,y,z=0).
\end{equation}
Here, $\nabla_\perp$ refers to the nabla operator in the $x$-$y$-plane, meaning $\nabla_\perp^2 = \partial^2/\partial x^2 + \partial^2/\partial y^2$ is the transverse Laplacian.
This formulation conserves the total intensity in the imaging plane $\Omega$ (assuming vanishing flux across its boundary region $\partial \Omega$) due to the divergence theorem
\begin{equation}
    -\frac{\delta \Delta}{\mu} \iint_{\Omega} \nabla_\perp \cdot \nabla_\perp I(x,y,z=0) \, \text{d}x \, \text{d}y = 0.
\end{equation}

The transmission projections used to obtain the X-ray attenuation coefficient volume in Fig.~\ref{fig:lead-rat-kidney} were corrected with a relative $\delta/\mu$-ratio based on the X-ray interaction properties of soft tissue and lead. An estimate for the soft tissue was obtained by averaging the electron number density and linear X-ray attenuation coefficients in an unstained piece of mouse kidney over a rectangular region of 100 consecutive slices; the values measured are $\rho_\text{e,1} = \SI{299}{\per\nano\metre\cubed}$ and $\mu_1 = \SI{0.567}{\per\centi\metre}$. The attenuation coefficient at an energy of \SI{25}{\kilo\eV} and the electron number density of atomic lead were obtained using a look-up table \cite{schoonjans2011xraylib} and are $\mu_2 = \SI{550}{\per\centi\metre}$ and $\rho_\text{e,2} = \SI{2399}{\per\nano\metre\cubed}$.
 
\subsection{Retrieval of electron number densities}
\label{sec:calibration}
The electron number density of the sample can be retrieved using X-ray phase measurements. UMPA tracks sub-pixel displacements $(u_x, u_y)$ of the reference pattern as a result of phase effects \cite{zdora2017x}. These displacements across the detector face in pixels can then be converted into refractive angles using
\begin{equation}
    \left(\phi_x, \phi_y\right) = \frac{2\pi}{\lambda} \,(u_x, u_y)\frac{p}{\Delta},
\end{equation}
where $(\phi_x, \phi_y)$ are the $x$- and $y$-derivatives of the object's phase $\phi(x,y)$, $\lambda$ is the wavelength, and $p$ is the pixel size at the detector face; note that the displacements $(u_x, u_y)$ are dimensionless because they have units of pixels \cite{zdora2017x}. The phase of the object can then be obtained by two-dimensional Fourier integration of the two differential phase signals \cite{frankot1988method, arnison2004linear, kottler2007two}:
\begin{equation}
    \phi(x, y) = \mathcal{F}^{-1}\left(\frac{\mathcal{F}(\phi_x + i \phi_y)}{ik_x - k_y}\right).
\end{equation}
It shall be noted that the Fourier space coordinates use the effective pixel size in the image plane (instead of at the detector face, hence now including magnification effects), although the distinction is not important if the calibration at the end of this section is performed. We define the phase of the air surrounding the object to be 0 by subtracting the average phase value in an air-filled region of interest.

The projected electron number density may be calculated using the recovered phase via
\begin{equation}
    \rho_{\text{e}, \perp}(x,y) = \frac{-\phi(x,y)}{\lambda\, r_e},
\end{equation}
where $r_\text{e}$ is the classical electron radius \cite{paganin2006coherent}. Finally, the electron number density in each voxel is obtained by tomographic reconstruction of these projections.

To minimize systematic errors in the obtained electron number density, we make use of a calibration procedure employed in Ref. \cite{zandarco2024speckle}. Here, a reference material is used to rescale the final position-dependent electron number density according to
\begin{equation}
\label{eq:calibration}
\rho_{e,\text{cal}}(x,y,z) = \frac{\rho_{e,\text{raw}}(x,y,z) - \bar{\rho}_{e,\text{bg}}}{\bar{\rho}_{e,\text{ref}} - \bar{\rho}_{e,\text{bg}}} \,\rho_{e,\text{theo}},
\end{equation}
where $\rho_{e,\text{cal}}(x,y,z)$ represents the three-dimensional calibrated electron number density map, $\rho_{e,\text{raw}}(x,y,z)$ is the electron number density map before calibration, $\bar{\rho}_{e,\text{bg}}$ and $\bar{\rho}_{e,\text{ref}}$ are the mean electron number density of the air background and reference material, respectively, and $\rho_{e,\text{theo}}$ is the theoretical electron number density of the reference material. In this work, we used a piece of poly(methyl methacrylate) (PMMA) as a reference material. Its theoretical electron number density is \SI{383}{\per\nano\metre\cubed}, which was calculated using the \emph{xraylib} library \cite{schoonjans2011xraylib}. Prior to analysis, all electron number density images were rescaled using Eq.~\eqref{eq:calibration}, setting the measured PMMA electron number density value to the theoretically expected value. The exception is the rat kidney sample from Sec.~\ref{sec:comparison-histology}, which was calibrated using its surrounding paraffin block as a reference material. The electron number density of the specific paraffin used in the experiment ($\rho_{e, \mathrm{paraffin}} = \SI{299}{\per\nano\metre\cubed}$) was known due to a computed tomography experiment for calibration that included a PMMA rod and paraffin.

\subsection{Material decomposition}
\label{sec:material-decomp}
Material decomposition aims to determine the composition of heterogeneous samples by expressing value pairs of complementary measurements in terms of known reference materials. The method makes use of the fact that any material with a linear attenuation coefficient $\mu$ and electron number density $\rho_\text{e}$ may be mathematically described as a linear combination of two base materials with attenuation coefficients and electron densities $\mu_i$ and $\rho_{\text{e}, i}$, respectively \cite{gureyev2015monomorphous}.
The volume fractions $v_i$ represent the relative amount of each base material contained in a voxel that reproduces both the measured attenuation coefficient and electron number density.
In order to obtain these volume fractions of the chosen base materials, a change of the basis of the coordinate system needs to be performed \cite{taphorn2022x}:
\begin{equation}
\begin{pmatrix} \mu \\ \rho_\text{e} \end{pmatrix} = \textbf{A}\, \begin{pmatrix} v_1 \\ v_2 \end{pmatrix} = \begin{pmatrix} \mu_1 & \mu_2 \\ \rho_{\text{e},1} & \rho_{\text{e},2} \end{pmatrix} \, \begin{pmatrix} v_1 \\ v_2 \end{pmatrix}.
\end{equation}
To obtain the material fractions, the vector of the obtained measurements $(\mu, \rho)$ is multiplied with the inverse change of basis matrix $\textbf{A}^{-1}$.

In Sec.~\ref{sec:virtual-histology}, a material decomposition was performed using soft tissue and lead as basis materials. Their respective X-ray interaction properties are provided at the end of Sec.~\ref{sec:edge-correction}. Because in Sec.~\ref{sec:comparison-histology} the attenuation value of native tissue at \SI{20}{\kilo\eV} was unknown, it was estimated using the lookup-table entry for water ($\mu_\text{w} = 0.81$).  Note that due to variations in the density of soft tissue, volume conservation (i.e. $v_1 + v_2 = 1$) does not necessarily have to be fulfilled \cite{taphorn2022x}. The volume fraction can be converted into a molar concentration
\begin{equation}
    c = v \frac{\rho}{M},
\end{equation}
provided the molar mass $M$ and mass density $\rho$ of the basis material are known.

\subsection{K-edge subtraction imaging}
\label{sec:k-edge-subtraction}
To provide validation of the modulation-based virtual histology method described in this paper, K-edge imaging of the same sample was performed. This approach can be used to retrieve the distribution of a target element by use of its K-edge, a sudden increase in X-ray absorption that occurs when the energy of the beam becomes sufficiently large to overcome the binding energy of the innermost electron shell of a material. An estimate of the volume fraction $v$ of a target material $X$ may be recovered by taking two complementary absorption tomography scans ($\mu_1, \mu_2$) at energies $E_K - \Delta E$ and $E_K + \Delta E$, where $E_K$ is the energy of the K-edge of the target material and $\Delta E$ is some small energy increment.

The attenuation in each voxel is attributable to a sum of the attenuation of the target material $X$ and all other materials, here referred to as the background $B$ \cite{meng2017model}:
\begin{equation}
    \mu_i = v\,\mu_\text{X}(E_K \pm \Delta E) + (1-v)\, \mu_\text{B}(E_K \pm \Delta E).
\end{equation}
Since any K-edges related to the background material are not within the energy scan range, which was chosen to cover the K-edge of the target material, we assume
\begin{equation}
    \mu_\text{B}(E_K + \Delta E) \approx \mu_\text{B}(E_K - \Delta E),
\end{equation}
allowing us to retrieve the approximate volume fraction of the target material using
\begin{equation}
\label{eq:k-edge-conversion}
    v_X = \frac{\mu_2 - \mu_1}{\mu_\text{X}(E_K + \Delta E) - \mu_\text{X}(E_K - \Delta E)}.
\end{equation}

\subsection{Comparison to optical histology}
\label{sec:comparison-optical-histology}
The histological slice was digitized using the ZEISS Axio Scan.Z1 microscope (\emph{Carl Zeiss AG}, Oberkochen, Germany) in conjunction with a 20-fold magnification objective. The image was converted to greyscale by computing the average of all three color channels and applying a percentile normalization in the range of \SI{1}{\percent} to \SI{99}{\percent}. Using the 2D-3D registration algorithm described in Ref.~\cite{chen20242d}, the 4 closest matching slices through the three-dimensional concentration volume were identified and deformably registered. Since a single slice in the reconstructed volume has a thickness of \SI{1.28}{\micro\metre}, the slice depicted in Fig.~\ref{fig:virtual-vs-real-histology}(b) shows a mean image of the 4 slices to match the slice thickness of conventional histology, which was \SI{5}{\micro\metre}. In addition, an unsharp mask as implemented in the \emph{scikit} python library \cite{scikit-image} (radius 1, strength 3) was applied to the image in panel (b).

The resolution in the virtual histology image was determined in the axial plane by performing a Fourier ring correlation \cite{harauz1986exact} in the implementation described in Ref.~\cite{riedel2023comparing}. The calculation was repeated for 100 consecutive slices, both using the half-bit and more conservative full-bit criterion, yielding the mean values stated in Sec.~\ref {sec:comparison-histology}.

\subsection{Offset rotation-axis scanning}
\label{sec:offset-rotation-axis}
\vspace{-1em}
A detector's horizontal field of view can be extended by up to a factor of 2 using offset rotation-axis scans. In this measurement protocol, the sample's rotation axis is offset such that its center lies closer to one edge of the detector (instead of the middle). The sample then needs to be imaged over a range of \SI{360}{\degree} instead of \SI{180}{\degree}. While the procedure is successfully used for propagation-based synchrotron imaging \cite{kawata2010microstructural, walsh2021imaging}, its application to modulation-based imaging data requires some additional image processing described in the following. 

After phase-retrieval with UMPA, each differential phase projection $\phi_x(x,y)$ and $\phi_y(x,y)$ is stitched with the projection recorded from the opposite angle (i.e.~180$^\circ$ apart), which is horizontally mirrored. In the case of $\phi_x(x,y)$, the sign of the mirrored projection needs to be flipped. As this process leads to redundant information in the overlap region, a linear blending is applied to provide a smooth transition -- as employed in Ref.~\cite{john2024centimeter}. The resulting differential phase projection of the full field of view is calculated using
\begin{equation}
\begin{split}
\phi_{i}(x,y) &= \alpha(x)\,\phi_{i, f}(x,y) \\
&\quad + (1-\alpha(x))\,\phi_{i, b}(x,y),
\end{split}
\end{equation}
where $i$ refers to either the differential phase signal in the $x$- or $y$-direction and $\phi_{i, b}$ are two differential phase projections spaced 180$^\circ$ apart. The linear blending function is defined as
\begin{equation}
\alpha(x) = \begin{cases}
1 & x < x_\mathrm{start} \\
1 - \frac{x - x_\text{start}}{\Delta w} & x_\mathrm{start} \leq x \leq x_\mathrm{start} + \Delta w \\
0 & x > x_\mathrm{start} + \Delta w.
\end{cases}
\end{equation}
Here, $\Delta w$ is the width of the overlap region and $x_{\text{start}}$ the $x$-coordinate corresponding to the start of the overlap. The transformation of the differential projections into electron densities then proceeds as previously described in Sec.~\ref{sec:calibration}.\\

\section*{Acknowledgments}
Parts of this research were carried out at PETRA III beamlines P05 and P07, operated by the Helmholtz-Zentrum Hereon. Beamtime was allocated under proposal IDs BAG-20240009 (P05) and BAG-20240010 (P07). We acknowledge DESY (Hamburg, Germany), a member of the Helmholtz Association HGF, for the provision of experimental facilities. This research was supported in part through the Maxwell computational resources operated at DESY. Part of this research was undertaken on the MCT beamline at the Australian Synchrotron, part of ANSTO. The authors gratefully acknowledge financial support by the ERC consolidator grant (Julia Herzen, TUM, DEPICT, PE3, 101125761) and by the ECI pathfinder (1MICRON, 101186826). Kaye S. Morgan acknowledges support from the Australian Research Council (FT18010037 and DP230101327). We thank the friendly staff of the Australian Synchrotron, Benedicta Arhatari, Andrew Stevenson, Adam Walsh, and Darren Thompson for their assistance. We also thank Linda Croton, Lorenzo D'Amico, Sophie-Marie Hornburg, James Pollock, and Stefan Schwaiger for useful discussions.

\section*{Author contributions}
D.J., J.He., and K.S.M. conceptualized the experiments. D.J., D.M.P., M.-C.Z., J.He. and K.S.M. developed the analysis methods. D.J., M.-C.Z., J.B.T., K.S.M., J.A., and S.J.A. prepared and performed the MCT experiment. L.M.P., P.I., S.B., and M.B. prepared the samples and optimized the protocols. J.C. performed the deformable registrations and matched the histology slice. L.M.P., P.I., J.A., and S.W. assisted with data analysis. D.J., F.B. and J.M. performed the experiment at P07. D.J. and J.U.H. performed the P05 experiment. All authors reviewed the manuscript.

\section*{Competing interests}
J.C. is an employee of ImFusion GmbH. All other authors declare no competing interests.

\section*{Data availability}
The datasets generated during and/or analyzed during the current study are available from the corresponding author on reasonable request.

\bibliography{bibliography}

\end{document}